\documentclass[aps,prl,twocolumn,superscriptaddress,showpacs,preprintnumbers,amsmath,amssymb]{revtex4}
\usepackage{graphicx} 
\usepackage{xspace}
\usepackage{epsfig}
\usepackage{multirow}
\usepackage{dcolumn}  
\usepackage[abs]{overpic}
\usepackage[section]{placeins}
\usepackage{color}

\newcommand{\de}{\ensuremath{\Delta E}\xspace}

\newcommand{\br}{\ensuremath{\mathcal{B}}\xspace}

\newcommand{\bb}{\ensuremath{B \overline{B}}\xspace}

\def\myspecial#1{}                   
\def\calL{{\mathcal L}}

\def\Mbc{M_{\rm bc}}
\def\etagg{\eta_{\gamma\gamma}}
\def\etapi{\eta_{3\pi}}
\def\Ebeam{E^{\,*}_{\rm beam}{}}
\def\vecpB{\vec{p}^{\,*}_B}
\def\vecpeta{\vec{p}^{\,*}_\eta}
\def\vecph{\vec{p}^{\,*}_{h}}
\def\EB{E_B^{\,*}{}}
\def\DeltaE{\Delta{E}}

\def\Eh{E^{\,*}_{h}}
\def\LR{{\mathcal R}}
\def\LRP{{\mathcal R^{\prime}}}
\def\pr{{Phys. Rev.}~}

\begin{document}

\myspecial{!userdict begin /bop-hook{gsave 300 50 translate 5 rotate
    /Times-Roman findfont 18 scalefont setfont
    0 0 moveto 0.70 setgray
    (\mySpecialText)
    show grestore}def end}



\preprint{\vbox{ \hbox{   }
						\hbox{Belle Preprint 2011-14}
                  \hbox{KEK Preprint 2011-16}
}}
\title{\quad\\[0.5cm]
Evidence for Direct CP Violation in $B^\pm \to \eta h^\pm$ and Observation of $B^0 \to \eta K^0 $}

\affiliation{Budker Institute of Nuclear Physics SB RAS and Novosibirsk State University, Novosibirsk 630090}
\affiliation{Faculty of Mathematics and Physics, Charles University, Prague}
\affiliation{University of Cincinnati, Cincinnati, Ohio 45221}
\affiliation{Gifu University, Gifu}
\affiliation{Hanyang University, Seoul}
\affiliation{University of Hawaii, Honolulu, Hawaii 96822}
\affiliation{High Energy Accelerator Research Organization (KEK), Tsukuba}
\affiliation{Indian Institute of Technology Guwahati, Guwahati}
\affiliation{Indian Institute of Technology Madras, Madras}
\affiliation{Institute of High Energy Physics, Chinese Academy of Sciences, Beijing}
\affiliation{Institute of High Energy Physics, Vienna}
\affiliation{Institute of High Energy Physics, Protvino}
\affiliation{Institute for Theoretical and Experimental Physics, Moscow}
\affiliation{J. Stefan Institute, Ljubljana}
\affiliation{Kanagawa University, Yokohama}
\affiliation{Institut f\"ur Experimentelle Kernphysik, Karlsruher Institut f\"ur Technologie, Karlsruhe}
\affiliation{Korea Institute of Science and Technology Information, Daejeon}
\affiliation{Korea University, Seoul}
\affiliation{Kyungpook National University, Taegu}
\affiliation{\'Ecole Polytechnique F\'ed\'erale de Lausanne (EPFL), Lausanne}
\affiliation{Faculty of Mathematics and Physics, University of Ljubljana, Ljubljana}
\affiliation{University of Maribor, Maribor}
\affiliation{Max-Planck-Institut f\"ur Physik, M\"unchen}
\affiliation{University of Melbourne, School of Physics, Victoria 3010}
\affiliation{Nagoya University, Nagoya}
\affiliation{Nara Women's University, Nara}
\affiliation{National Central University, Chung-li}
\affiliation{National United University, Miao Li}
\affiliation{Department of Physics, National Taiwan University, Taipei}
\affiliation{H. Niewodniczanski Institute of Nuclear Physics, Krakow}
\affiliation{Nippon Dental University, Niigata}
\affiliation{Niigata University, Niigata}
\affiliation{Osaka City University, Osaka}
\affiliation{Pacific Northwest National Laboratory, Richland, Washington 99352}
\affiliation{Panjab University, Chandigarh}
\affiliation{Research Center for Nuclear Physics, Osaka}
\affiliation{Seoul National University, Seoul}
\affiliation{Sungkyunkwan University, Suwon}
\affiliation{School of Physics, University of Sydney, NSW 2006}
\affiliation{Tata Institute of Fundamental Research, Mumbai}
\affiliation{Excellence Cluster Universe, Technische Universit\"at M\"unchen, Garching}
\affiliation{Tohoku Gakuin University, Tagajo}
\affiliation{Tohoku University, Sendai}
\affiliation{Department of Physics, University of Tokyo, Tokyo}
\affiliation{Tokyo Institute of Technology, Tokyo}
\affiliation{Tokyo Metropolitan University, Tokyo}
\affiliation{Tokyo University of Agriculture and Technology, Tokyo}
\affiliation{CNP, Virginia Polytechnic Institute and State University, Blacksburg, Virginia 24061}
\affiliation{Yonsei University, Seoul}
  \author{C.-T.~Hoi}\affiliation{Department of Physics, National Taiwan University, Taipei} 
  \author{P.~Chang}\affiliation{Department of Physics, National Taiwan University, Taipei} 
  \author{H.~Aihara}\affiliation{Department of Physics, University of Tokyo, Tokyo} 
  \author{D.~M.~Asner}\affiliation{Pacific Northwest National Laboratory, Richland, Washington 99352} 
  \author{T.~Aushev}\affiliation{Institute for Theoretical and Experimental Physics, Moscow} 
  \author{A.~M.~Bakich}\affiliation{School of Physics, University of Sydney, NSW 2006} 
  \author{K.~Belous}\affiliation{Institute of High Energy Physics, Protvino} 
  \author{V.~Bhardwaj}\affiliation{Panjab University, Chandigarh} 
  \author{B.~Bhuyan}\affiliation{Indian Institute of Technology Guwahati, Guwahati} 
  \author{M.~Bischofberger}\affiliation{Nara Women's University, Nara} 
% ...
\collaboration{The Belle Collaboration}

\begin{abstract}
 We report measurements of the branching fractions and $CP$
 asymmetries for $B^\pm \to \eta h^\pm$ ($h = K$ or $\pi$) and the observation of
 the decay  $B^0 \to\eta K^0$ from the final data sample of 
 772~$\times~10^6$ $B\overline{B}$ pairs collected with the Belle
detector at the
KEKB asymmetric-energy $e^+e^-$ collider. The measured branching fractions are
$\br(B^{\pm}\to \eta K^{\pm}) =
(2.12 \pm 0.23 \pm 0.11)\times 10^{-6}$, $\br(B^{\pm}\to \eta \pi^{\pm}) =
(4.07 \pm 0.26 \pm 0.21 )\times 10^{-6}$
and  $\br(B^0\to
\eta K^0) = (1.27^{+0.33}_{-0.29} \pm 0.08)\times
10^{-6}$, where the last decay is observed for the first time with a significance of 5.4 standard
deviations ($\sigma$). We also find evidence for $CP$ violation
in the
charged $B$ modes, $A_{CP}(B^\pm \to\eta K^\pm) = -0.38 \pm 0.11 \pm 0.01$ and
$A_{CP}(B^\pm \to\eta \pi^\pm) = -0.19\pm 0.06 \pm 0.01$ with significances of
$3.8\sigma$ and $3.0\sigma$, respectively. For all measurements,
 the first and second uncertainties are
statistical and systematic, respectively.


\end{abstract}


\maketitle

\tighten

{\renewcommand{\thefootnote}{\fnsymbol{footnote}}
\setcounter{footnote}{0}




Charge-Parity ($CP$) violation plays an important role in any explanation of the observed
 dominance of matter over antimatter in our Universe. Current experimental knowledge about $CP$ violation is limited.
 Charmless hadronic $B$ decays constitute sensitive probes for $CP$ violation in the standard model (SM) as well as beyond,
 and can help to further elucidate this unsolved question. In the SM,
the decays $B^\pm\to \eta^{(\prime)} K^\pm$ and $B^0 \to \eta^{(\prime)} K^0$ are expected to  
primarily proceed through $b\to s$ penguin processes and a 
$b\to u$ tree transition as shown in Fig.~\ref{fig:ksfeynman}. 
 The large $B\to \eta^\prime K$\cite{etaphcleo,etaphbelle, etaphbabar}
and small $B\to \eta K$\cite{etahbelle1,etaphbabar} branching fractions can be explained
by $\eta-\eta^\prime$ mixing along with constructive and destructive interference  
between  the amplitudes of the two penguin processes \cite{lipkin}.

The branching fraction of $B^0 \to \eta K^0$ is expected to be lower than that of
 $B^\pm \to \eta K^\pm$, because the tree diagram in the $B^0 \to \eta K^0$ decay is color suppressed. 
The destructive combination of penguin amplitudes may interfere with the tree amplitude in $B \to \eta K$,
resulting in a large direct $CP$ asymmetry ($A_{CP}$) \cite{lipkin,soni}, defined as 
\begin{equation}
 A_{CP} \equiv \frac{\Gamma[B^- (\overline{B}{}^0)\to \eta h^{- (0)}] - \Gamma[B^+ (B^0) \to \eta h^{+ (0)}]}
          {\Gamma[B^- (\overline{B}{}^0) \to \eta h^{- (0)}] +\Gamma[B^+ (B^0)\to \eta h^{+ (0)}]},
\end{equation}
where $\Gamma(B\to\eta h)$ is the partial width obtained for the $B\to\eta h$ decay, and $h$ denotes $K$ or $\pi$. 
Similarly, direct $CP$ violation could be sizeable for $B^\pm\to \eta \pi^\pm$ owing to 
the interference between $b\to d$ penguin and $b\to u$ tree diagrams. Several theoretical calculations with different mechanisms \cite{revisit,XiaoKeta,Xiaopieta,ChenKeta,Zupan,MBMN} 
 suggest a large $A_{CP}$ for both $B\to\eta K$ and $B\to\eta \pi$,  
although the sign could be either positive or negative. Previous Belle 
\cite{etahbelle1} and BaBar \cite{etaphbabar} measurements indicate a large negative $A_{CP}$ 
in $B^\pm\to \eta K^\pm$, but more data are needed to be statistically sensitive to a non-zero $A_{CP}$ in $B^\pm\to \eta \pi^\pm$.

\begin{figure}[b!]
\includegraphics[width=0.5\textwidth]{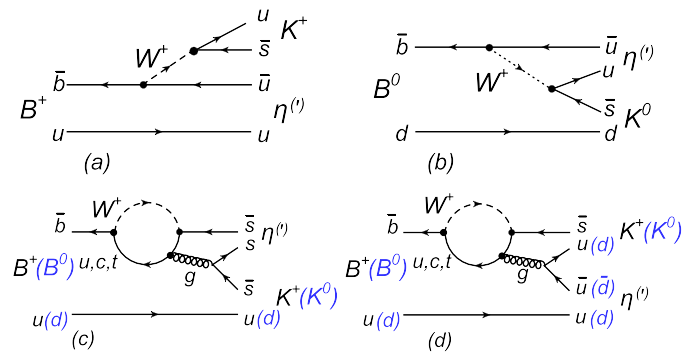}
\caption{(a) $b\to u$ tree diagram for $B^+ \to\eta^{(\prime)} K^+$. (b) color-suppressed $b\to u$ tree diagram for $B^0 \to\eta^{(\prime)} K^0$.
 (c), (d) $ b\to s$
gluonic penguin diagrams for $B \to\eta^{(\prime)} K$.}
\label{fig:ksfeynman}
\end{figure}

In this Letter, we report the first observation of $B^0 \to\eta K^0$, and 
evidence for direct $CP$ asymmetries in $B^\pm \to \eta K^\pm$ and $B^\pm\to \eta \pi^\pm$
using the final Belle data set.
The data sample corresponds to (772 $\pm$ 11)~$\times~10^6$
$\bb$ pairs collected with the Belle detector at the KEKB $e^+e^-$ asymmetric-energy
(3.5~GeV on 8.0~GeV) collider~\cite{kur} operating at the $\Upsilon(4S)$ resonance. 
The production rates of $B^+B^-$ and $B^0\overline{B}{}^0$ pairs are assumed to
be equal at the $\Upsilon(4S)$ resonance.

The Belle detector ~\cite{aba} is a large-solid-angle magnetic
spectrometer that consists of a silicon vertex detector,
a 50-layer central drift chamber (CDC), an array of
aerogel threshold Cherenkov counters (ACC),
a barrel-like arrangement of time-of-flight
scintillation counters (TOF), and an electromagnetic calorimeter
comprising CsI(Tl) crystals located inside
a superconducting solenoid coil that provides a 1.5~T
magnetic field.  An iron flux-return located outside
the coil is instrumented to detect $K_L^0$ mesons and to identify
muons.

The event selection and $B$ candidate reconstruction methods are similar to 
those described in Ref. \cite{etahbelle1}. 
We select $\eta$ and $\pi^0$ candidates through the decay chains
$\eta\to \gamma\gamma$ ($\eta_{\gamma\gamma}$), $\eta\to \pi^+\pi^-\pi^0$
 ($\etapi$) and $\pi^0 \to \gamma \gamma$. We require the two photons from the $\eta$ and $\pi^0$ decays to have 
energies $(E_{\gamma i},~i=1,2)$
greater than 50 MeV. Candidate $\pi^0$ mesons are selected from pairs of photons
with invariant masses between 115 MeV/$c^2$ and
152 MeV/$c^2$. 
In the $\etagg$ reconstruction, the 
photon energy asymmetry $|E_{\gamma 1}-E_{\gamma 2}|/(
   E_{\gamma 1}+E_{\gamma 2})$, is required to be  
less than 0.9 to reduce the large combinatorial background from low-energy photons.
Photons in $\eta_{\gamma\gamma}$ reconstruction 
are not allowed
to pair with any other photon having $E_{\gamma}>100~\mathrm{MeV}$, to form a 
$\pi^0$
candidate.
We require the invariant mass of the $\eta_{\gamma\gamma}$ and $\eta_{3\pi}$ candidates to be in the
intervals $(501,573)~\mathrm{MeV}/c^2$ and $(538.5,556.5)~\mathrm{MeV}/c^{2}$, respectively. 
In order to improve the $\pi^0$ and $\eta$ energy resolution, a mass-constrained kinematic fit
is performed after the candidate selection.

Charged tracks that are not used to form
$K^{0}_{S}$ candidates (see below) are required to have a distance of closest approach with respect
to the interaction point (IP) of less than 3.0~cm along the electron beam
direction ($z$) and less than 0.3 cm in  the transverse plane.
Charged kaons and pions are identified using $dE/dx$ information
from the CDC, Cherenkov light yields in the ACC, and time-of-flight information from the TOF. This information is
combined to form a likelihood ratio, $\mathcal{R}_{K/\pi}
= \mathcal{L}_K/(\mathcal{L}_K+\mathcal{L}_\pi)$, where
$\mathcal{L}_{K}$ $(\mathcal{L}_{\pi})$ is the likelihood of the
track being a kaon (pion). Charged tracks with
$\mathcal{R}_{K/\pi}>0.6$ ($<0.4$) are treated as kaons (pions) for $B^{\pm}\to
\eta K^{\pm}$ ($B^{\pm} \to \eta \pi^{\pm}$) reconstruction. 
A less stringent requirement,
${\mathcal R}_{K/\pi} <0.6$, is used for charged pions in the $\eta_{3\pi}$
selection. The efficiencies for the $\mathcal{R}_{K/\pi}$ requirement are
$84\%$ for kaons, $89\%$ for pions, and $94\%$ for pions in $\etapi$ reconstruction. Furthermore, for
$B^{\pm} \to \eta h^{\pm}$ and $B^{0} \to \eta K^{0}$ we reject charged tracks consistent with either the 
electron or muon hypothesis.

Candidate $K^0$ mesons are reconstructed in $K^0_S \to \pi^+ \pi^-$
decays. The $K^0_S$ candidates 
are required to have an invariant mass lying between 488 MeV/$c^2$ and 508 MeV/$c^2$.
The charged tracks for each $K^0_S$ candidate are required to have a distance-of-closest approach with respect
to the IP of larger than 0.02~cm in the transverse plane. The 
angle between the $K^0_S$ momentum
 and the direction from the IP to the $K^0_S$ decay vertex must be within 0.03 rad. The distance between the two
daughter tracks at their point of closest approach in the transverse plane is required to be less than 2.40 cm, 
and the flight length of the $K^0_S$ is required 
to be larger than 0.22 cm.

Candidate $B$ mesons are identified using the modified beam-energy-constrained mass 
 $\Mbc=\sqrt{(\Ebeam/c^2)^2 - |\vecpB/c|^2}$ \cite{kstarg},
and the energy difference $\DeltaE=\EB - \Ebeam$,
where $\Ebeam$ is the
beam energy, and $\EB$ and $\vecpB$ are the energy and modified momentum,
respectively, of the $B$ candidate in the $\Upsilon(4S)$ rest frame.  The energy $\EB$ is
calculated as $E^{*}_{B}=E^{*}_{\eta}+E^{*}_{h}$. 
The momentum $\vecpB$ is calculated
according to
\begin{equation}
\vecpB=\vecph + {\vecpeta\over|\vecpeta|} \times \sqrt{(\Ebeam - \Eh)^2-m_{\eta}^2},
\end{equation}
where $m_{\eta}$ is the nominal $\eta$ mass \cite{PDG}. Since charged tracks are generally
measured with a better precision than photons, the 
$\eta$ decays to neutral particles have worse momentum resolution than primary
charged tracks from $B$ decays. The $\vec{p}_{B}^{*}$
resolution is improved using Eq. (2) because $\vec{p}^*_h$ and $E_{\mathrm{beam}}^{*}$ are determined more precisely
than $\vec{p}^{*}_{\eta}$.
Events with $\Mbc >5.2$ GeV/$c^2$ and $|\de|<0.3$ GeV are retained for further 
 analysis.

The dominant background arises from $e^+e^- \to q\overline q ~(
q=u,d,s,c )$ continuum events. We use event topology variables to
distinguish spherically distributed $B\overline{B}$
events from the jet-like continuum background. First we combine a set of
modified Fox-Wolfram moments \cite{pi0pi0} into a Fisher
discriminant. We then compute a likelihood from the product of probability density functions (PDFs) that describe the Fisher discriminant, $\cos{\theta^*_B}$, and $\Delta z$ distributions. Here, $\theta^*_B$ is 
the angle between the $B$ flight direction and the beam direction
in the $\Upsilon(4S)$ rest frame, and $\Delta z$ is the decay flight-length difference, along the $z$ axis, between vertices of the signal $B$ and the accompanying $\overline{B}$. A likelihood ratio, 
$\LR = {\calL}_s/({\calL}_s + {\calL}_{q \bar{q}})$, is formed from signal
(${\calL}_s$) and background (${\calL}_{q \bar{q}}$) likelihoods, which are obtained 
from GEANT-based~\cite{geant} Monte Carlo (MC) simulated samples.
Signal MC events are generated with EVTGEN \cite{EVTGEN}, which invokes the PHOTOS \cite{photos} package  to take 
final state radiation into account. 
We require $\mathcal{R}>0.2$ to suppress continuum background in all modes and then translate $\LR$ to $\LRP$, defined as
 \begin{eqnarray}
\LRP=\ln\left(\frac{\LR-\LR_{\mathrm{min}}}{\LR_{\mathrm{max}}-\LR}\right)  \mbox{.}
\end{eqnarray}
In this expression $ \LR_{\mathrm{min}}$ ($\LR_{\mathrm{max}}$) is equal to 0.2 (1.0). 
This translation is convenient, as the $\mathcal{R}^{\prime}$
distributions for signal and backgrounds can be described by a simple sum of Gaussian
functions.

\begin{table*}[htb]
\begin{center}
\caption{
Detection efficiency ($\epsilon$) including sub-decay branching fractions, yield, significance of the yield $\Sigma(\cal{Y})$, measured branching fraction $\cal{B}$, charge asymmetry $A_{CP}$, and significance of the charge asymmetry $\Sigma(A_{CP})$ for $B\to \eta h$ decays.
The first errors are statistical and the second ones are systematic.}
\begin{tabular}{c|ccccccc}
\hline\hline
 Mode                         & $\epsilon$ (\%) & Yield                     &$\Sigma\cal{(Y)}$& $\cal{B}$ $(10^{-6})$ & $A_{CP}$  & $\Sigma(A_{CP})$ \\
\hline      
$B^\pm \to\eta K^\pm$          &               &                             & 13.2   &      $2.12 \pm 0.23 \pm 0.11$   &  $-0.38 \pm 0.11 \pm 0.01$     &  3.8   \\
$\eta_{\gamma \gamma} K^\pm$   & 13.3        &        $201.9^{+27.1}_{-26.5}$       & 10.2 &  $2.07 \pm 0.27 \pm 0.10$ & $-0.36\pm{0.13}\pm 0.01$  & 2.9 \\
$ \eta_{3\pi} K^\pm$          & 4.9         &   $80.2^{+14.9}_{-13.9}$      & 8.6  &   $2.29^{+0.43}_{-0.40} \pm 0.15$   & $-0.42 \pm 0.18\pm 0.01$ & 2.4 \\
\hline
$B^\pm \to\eta \pi^\pm$        &        &                                    &  22.4   &      $4.07 \pm 0.26 \pm 0.21$    &$-0.19\pm 0.06\pm 0.01$  &3.0 \\
$\eta_{\gamma \gamma} \pi^\pm$ & 15.3 &   $480.6^{+35.1}_{-36.0}$ &     19.0   &   $4.24 \pm 0.32 \pm 0.19$ & $-0.14 \pm 0.08\pm 0.01$  &1.8 \\
$ \eta_{3\pi} \pi^\pm$         & 5.4  & $138.6^{+18.5}_{-17.5}$         &  12.2  &  $3.63\pm 0.49 \pm 0.25$ &$-0.31  \pm 0.13 \pm 0.01$  &2.5 \\
\hline
$B^0 \to\eta K^0$              &        &                                    &   5.4    &          $1.27^{+0.33}_{-0.29} \pm 0.08$           & \\
$\eta_{\gamma \gamma} K^0$     &    4.2 &   $38.0^{+12.6}_{-11.4}$   &  4.0 &    $1.18^{+0.39}_{-0.35}\pm 0.06$   &  \\
$ \eta_{3\pi} K^0$              & 1.5    &   $16.2^{+6.5}_{-5.4}$     &  4.1  &   $1.48^{+0.59}_{-0.49} \pm 0.10$    &   \\
\hline\hline
\end{tabular}
\label{tab:etah}
\end{center}
\end{table*}

\begin{figure*}
\begin{overpic}[totalheight=34mm,width=0.9\textwidth]
{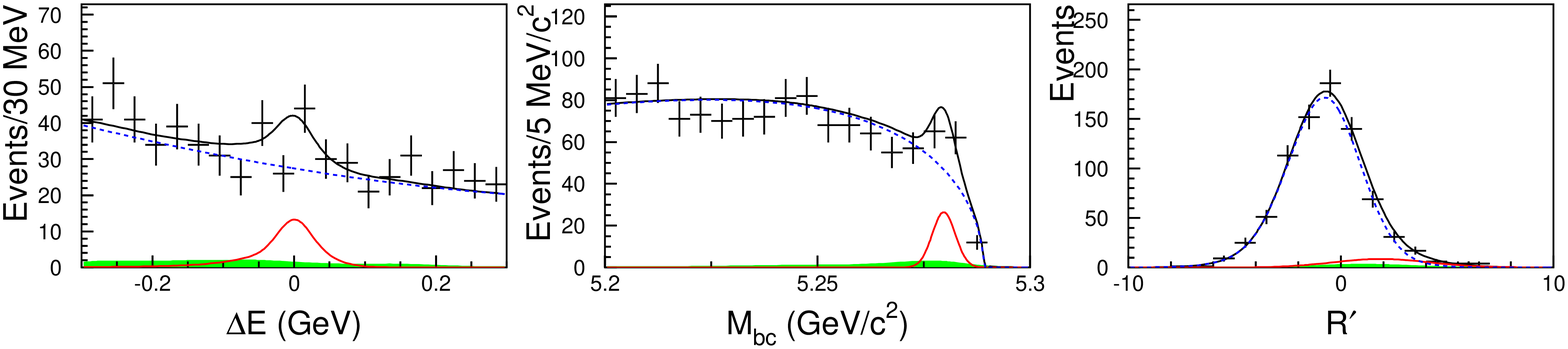}
\put(30,83){$B^0\to\eta K^0$}
\put(185,83){$B^0\to\eta K^0$}
\put(335,83){$B^0\to\eta K^0$}
\end{overpic}
\caption{$\Delta E$ (left), $M_{\rm bc}$ (middle) and $\LRP$ (right) distributions
for $B^0 \to \eta K^0$ candidate events with 
the $\etagg$ and $\etapi$ modes combined. Points with error bars represent the data,
the total fit functions are shown by black solid curves, signals are shown by red solid curves, dashed curves show 
the continuum contributions and filled histograms are the contributions
from charmless $B$ backgrounds. The $\de$, $\Mbc$, and $\LRP$ projections of the fit are for events that have
 $5.27$ GeV/$c^2 < {M_{\rm bc}} <5.30$ GeV/$c^2$ and $\LRP >$0.55, 
$-0.10$ GeV$ < \de <0.08$ GeV and $\LRP >$0.55,
and $-0.10$ GeV$ < \de <0.08$ GeV and  $5.27$ GeV/$c^2 < {M_{\rm bc}} <5.30$ GeV/$c^2$, respectively.
} 
\label{fig:ks}
\end{figure*}

Signal yields are extracted by performing an unbinned extended three-dimensional
maximum likelihood fit.
The likelihood for each $B^+$ mode is defined as
\begin{eqnarray} 
 \mathcal{L} & = &  e^{ -\sum_{j} N_j}
\times \prod_i (\sum_j N_j \mathcal{P}^i_j) \mbox{,}  \nonumber\\ 
\mathcal{P}^i_j & = &\frac{1}{2}[1- q^i  A_{CPj}]
P_j(M^i_{\rm bc}, \Delta E^i, \LRP^i),
\end{eqnarray}
where $i$ denotes the $i^{\mathrm{th}}$ event. Category $j$ runs over the signal and background components, 
where  background components include continuum,
the cross-feed due to $K$-$\pi$ misidentification, and the
background from other charmless $B$ decays, but do not include $b \to c$ decays, which are found to give a negligible contribution.
The parameter $N_j$ represents the number of events for category $j$, $P_j(M_{\rm bc}^i, \Delta E^i, \LRP^i)$ is the PDF in
$M_{\rm bc}$, $\Delta E$ and $\LRP$, and $q$ is the $B$-meson charge.
For the $B^0$ mode, the $A_{CPj}$
parameters are set to zero since
$CP$ asymmetries cannot be determined
without additional information.
 The validity of the three-dimensional fit is tested
with large ensemble tests using MC samples and by fits to a high-statistics control sample of $B^{+}\to
\overline{D}{}^0 \pi^{+}(\overline{D}{}^0 \to K^{+}\pi^{-}\pi^{0})$ decays.

All the signal and cross-feed background PDFs in $M_{\rm bc}$ and $\LRP$ are modeled with a single Gaussian.
In $B \to \etagg h$ ($B \to \etapi h$) modes, the PDFs in $\Delta E$ are described by a Crystal Ball \cite{cbf}
(a sum of two Gaussians) function .
The peak positions and resolutions in $\Mbc$, $\de$ and $\LRP$
are adjusted according to the data-MC differences observed in control samples \cite{control}.

The continuum background in $\de$ is described by a second-order
polynomial, while the $\Mbc$ distribution is parameterized with an 
ARGUS function, $f(x) = x \sqrt{1-x^2}\;{\rm exp}\;[ -\xi (1-x^2)]$, where 
$x$ is $M_{\mathrm{bc}}/E_{\mathrm{beam}}$ and $\xi$ is a free parameter in the fit \cite{argus}. The
$\mathcal{R}^{\prime}$ PDF is a double Gaussian function.
The background PDFs in both $\Mbc$ and $\de$ for charmless $B$ decays are modeled with
smoothed two-dimensional histograms obtained from a large MC sample, and the $\mathcal{R}^{\prime}$ PDF is a single Gaussian.

We perform a simultaneous fit
to $B^{\pm} \to \eta K^{\pm}$ and $B^{\pm} \to \eta \pi^{\pm}$ candidate events, since
these two decay modes can feed into each other. In the likelihood fits,
$N_{j}$ and $A_{CPj}$ 
are allowed to vary for the continuum and charmless $B$ backgrounds.
Shape parameters of the continuum background PDFs are also floated. 
The cross-feed background in $\eta K^{\pm}$ $(\eta\pi^{\pm})$ and signal in $\eta \pi^{\pm}$ $(\eta K^{\pm})$
share the same fitting parameters in both $A_{CP}$ and branching fraction. 
Figure~\ref{fig:ks} shows the $M_{\rm bc}$, $\Delta E$ and $\LRP$ projections of the fit to the $B^0 \to \eta K^0_{S}$ sample.
The $M_{\rm bc}$ and $\Delta E$ projections for the $B^+\to \eta h^+$ and $B^-\to \eta h^-$ samples are shown separately in Fig.~\ref{fig:kpi}.

The branching fraction for each mode is calculated by dividing the 
efficiency-corrected signal yield by the number of $B\overline{B}$
pairs. The dominant systematic errors on the branching fraction come from MC modeling of the $\eta$,
$\pi^{0}$, and $K^{0}_{S}$ selection efficiency; these errors are 
$4.0\%$, $4.0\%$, and $1.6\%$, respectively. The systematic error 
due to ${\cal R}(K/\pi)$ requirement is $0.9\%$ for kaons and $0.8\%$ for pions. It is estimated from the $D^{*+}\to D^{0}\pi^{+}(D^0 \to K^{-}\pi^{+})$ sample.
The systematic error due to the charged-track reconstruction efficiency is estimated
to be $0.35$\% per track, which is determined from a study of the $D^{*\pm}\to D^0 \pi^{\pm}(D^0 \to \pi^+ \pi^- K^0_{S})$ decay. Any difference in the efficiency when the $\mathcal{R}$ criterion is applied to data or MC is investigated using the $B^{+}\to\overline{D}{}^0\pi^{+}(\overline{D}{}^0 \to K^{+}\pi^{-}\pi^{0})$ sample; the results of this study imply a 0.6\% systematic uncertainty.   
The fitting systematic errors due to the signal PDF modeling
are estimated from changes of the fit parameters while varying the calibration factors by one standard
deviation. 
The systematic error due to charmless $B$ background PDF modeling is calculated from the difference observed 
between the signal yield when the charmless yield is floated in the fit and that when the yield is fixed to the MC expectation.
The systematic error due to the uncertainty in the total number of $\bb$ pairs 
is 1.4\%, and the error due to limited signal MC statistics used to evaluate the efficiency is 0.55\%.
 The systematic errors on $A_{CP}$ arise from detector bias, uncertainties on the detector bias
 and PDF modeling. 
The possible detector bias due to the tracking acceptance and ${\cal R}(K/\pi)$ selection   
for $A_{CP}(B^\pm\to \eta \pi^\pm)$ is evaluated using the fitted $A_{CP}$ value of the continuum background \cite{kbias1,kbias2}. 
The detector bias in $A_{CP}(B^\pm\to \eta K^\pm)$ is evaluated using the $D^+_s \to \phi \pi^+ (\phi \to K^+ K^-)$ and
$D^0 \to K^- \pi^+$ samples \cite{kbias1,kbias2}.
There is a contribution to the $A_{CP}$
systematic uncertainty from the modeling of the signal PDFs.
The total systematic errors for $A_{CP}$ are in the range $(8.2 - 14.2)\times 10^{-3}$. 

\FloatBarrier
 The statistical significance
is evaluated as $\sqrt{-2\ln(\calL_0/\calL_{\rm max})}$, where ${\cal L}_0$ is the likelihood value when either the signal yield or $A_{CP}$ is fixed to zero, and $\calL_{\rm max}$ is the nominal likelihood value. The total significance  (${\Sigma}$)
including PDF modeling systematic uncertainty is calculated after smearing 
the likelihood distribution with the appropriate PDF modeling systematic error.
In Table \ref{tab:etah} we list the fitted signal yields,
charge asymmetries, reconstruction efficiencies,
and branching fractions. The combined result for the two $\eta$ decay modes is obtained from the combined likelihood function. 
\begin{figure}[h!]
\begin{overpic}[width=88mm]
{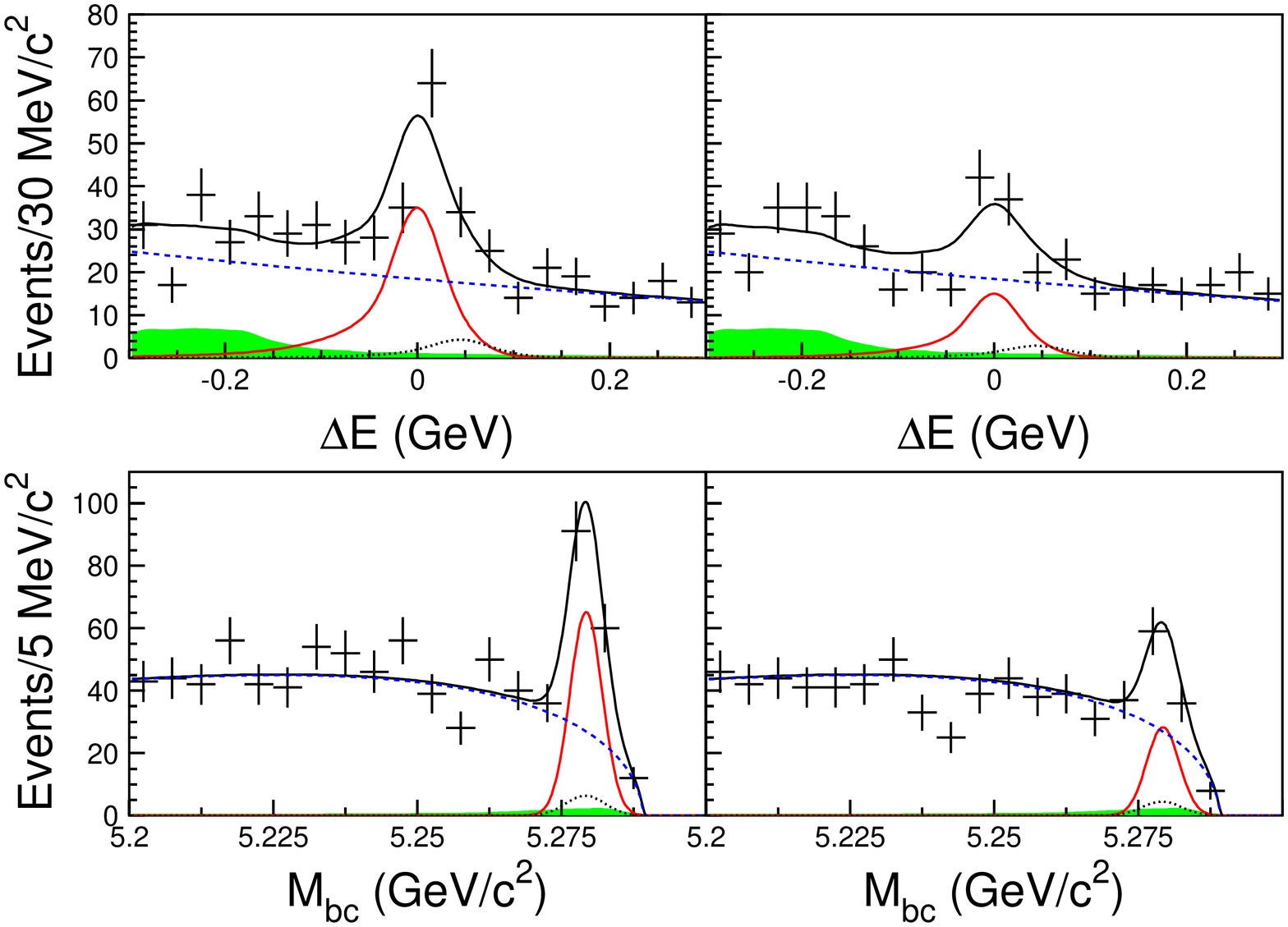}
\put(29,168){$B^+\to\eta K^+$}
\put(141,168){$B^-\to\eta K^-$}
\put(29,78){$B^+\to\eta K^+$}
\put(141,78){$B^-\to\eta K^-$}
\end{overpic}
\hspace{-0.4cm}

\begin{overpic}[width=88mm]
{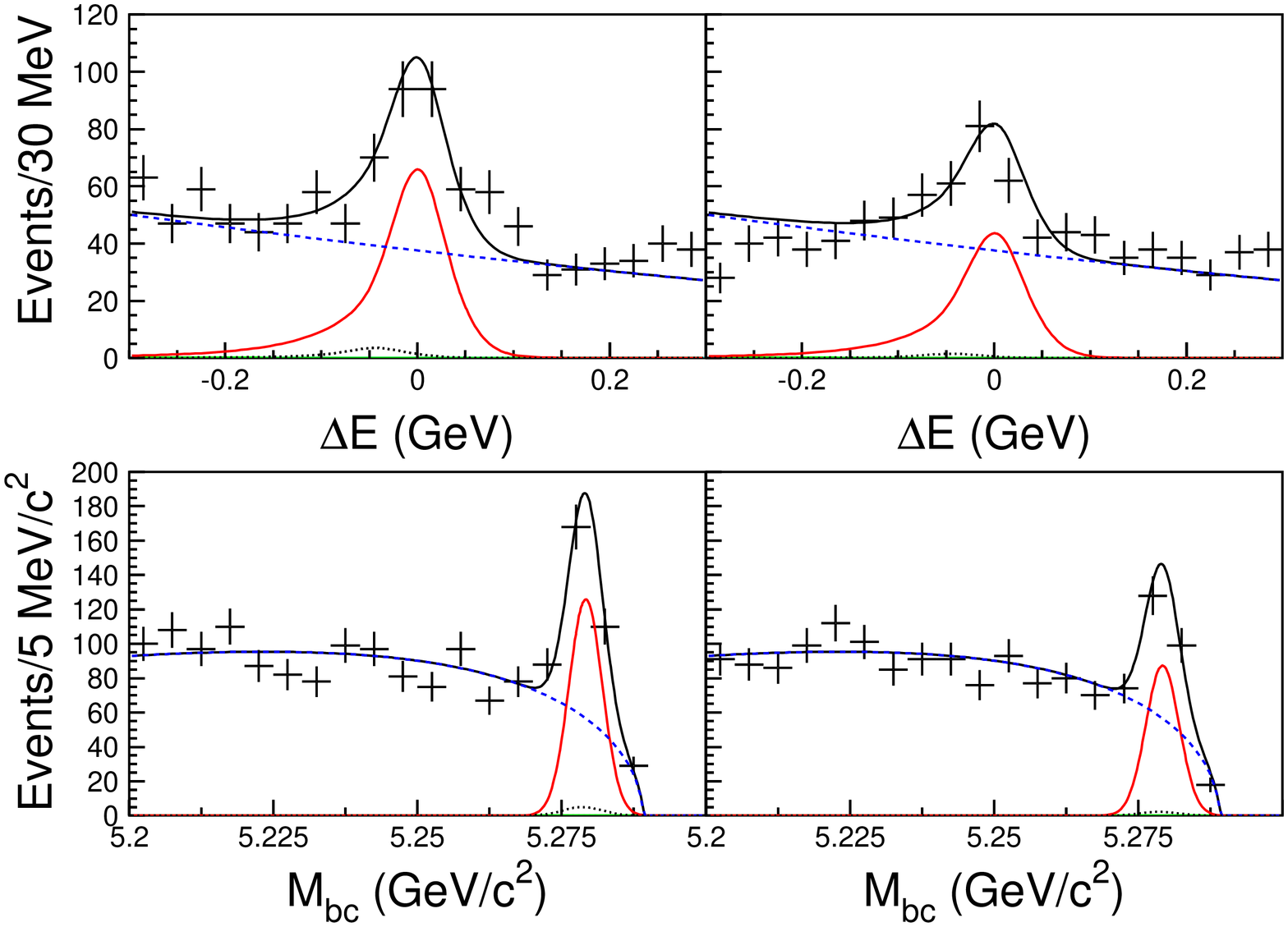}
\put(29,168){$B^+\to\eta \pi^+$}
\put(141,168){$B^-\to\eta \pi^-$}
\put(29,78){$B^+\to\eta \pi^+$}
\put(141,78){$B^-\to\eta \pi^-$}
\end{overpic}

\caption{$\de$ and $\Mbc$ projections for
 $B^+ \to \eta h^+$ (left) and $B^- \to \eta h^-$ (right) candidate events with 
the $\etagg$ and $\etapi$ modes combined. Points with error bars represent the data,
the total fit functions are shown by black solid curves, signals are shown by red solid curves, dashed curves are 
the continuum contributions, dotted curves are the cross-feed backgrounds from misidentification
and filled histograms are the contributions
from charmless $B$ backgrounds. The $\de$ and $\Mbc$ projections of the fits are for events that have
 $5.27$ GeV/$c^2 < {M_{\rm bc}} <5.30$ GeV/$c^2$ and $\LRP >$1.95, 
$-0.10$ GeV$ <\de <0.08$ GeV and $\LRP >$1.95, respectively.
}
\label{fig:kpi}
\end{figure}

In summary, using the final Belle data sample containing 772~$\times~10^6$ $B\overline{B}$ pairs and a three-dimensional fit that maximizes the efficiency,
we provide new measurements based on signal yields 2.5 times larger than those reported in our previous publications  \cite{etahbelle1}. 
We find evidence for $CP$ asymmetries in $B^{\pm}\to \eta K^{\pm}$ and $B^{\pm}\to \eta \pi^{\pm}$: $A_{CP}(B^\pm \to\eta K^\pm) = -0.38 \pm 0.11 \pm 0.01$ and
$A_{CP}(B^\pm \to\eta \pi^\pm) = -0.19\pm 0.06 \pm 0.01$. The significance of $A_{CP}(\eta K^+)$ $[A_{CP}(\eta \pi^+)]$ is $3.8\sigma$ $[3.0\sigma]$.
Evidence for $A_{CP}(B^\pm \to\eta \pi^\pm)$ is seen for the first time.
We also observe the decay $B^0 \to\eta K^0$
for the first time with a significance of $5.4\sigma$ and a branching fraction $\br(B^0\to
\eta K^0) = (1.27^{+0.33}_{-0.29} \pm 0.08)\times
10^{-6}$.
In addition, we report the following new measurements of the branching
fractions: $\br(B^{\pm}\to \eta K^{\pm}) =
(2.12 \pm 0.23 \pm 0.11)\times 10^{-6}$ and $\br(B^{\pm}\to \eta \pi^{\pm}) =
(4.07 \pm 0.26 \pm 0.21 )\times 10^{-6}$. All our branching
fraction and $A_{CP}$ measurements supersede the results in Ref.\cite{etahbelle1}.

We thank the KEKB group for excellent operation of the
accelerator, the KEK cryogenics group for efficient solenoid
operations, and the KEK computer group and
the NII for valuable computing and SINET4 network support.  
We acknowledge support from MEXT, JSPS and Nagoya's TLPRC (Japan);
ARC and DIISR (Australia); NSFC (China); MSMT (Czechia);
DST (India); MEST, NRF, NSDC of KISTI, and WCU (Korea); MNiSW (Poland); 
MES and RFAAE (Russia); ARRS (Slovenia); SNSF (Switzerland); 
NSC and MOE (Taiwan); and DOE and NSF (USA).



\begin{thebibliography}{99}
\bibitem{control} The $\Delta E$ peak position in the $B\to \etagg h$ [$B\to \etapi h$] decay is calibrated using 
the $B^{+}\to \pi^{0} K^+$ [$B^{+}\to\overline{D}{}^0\pi^{+}(\overline{D}{}^0 \to K^{+}\pi^{-}\pi^{0})$] sample.
The resolution in $\de$ is adjusted using the $\overline{D}{}^0 \to K^{+}\pi^{-}\pi^{0}$ samples.
The peak positions and resolutions in $\Mbc$ and $\LRP$
are adjusted using the $B^{+}\to\overline{D}{}^0\pi^{+}(\overline{D}{}^0 \to K^{+}\pi^{-}\pi^{0})$ control sample.
\end{thebibliography}
\end{document}